\documentclass[reqno]{amsart}
\usepackage{amsmath}
\usepackage{amssymb}
\usepackage{amsfonts}
\usepackage{mathrsfs}
\usepackage[all,arc]{xy}
\usepackage{enumerate}
\usepackage{mathrsfs}
\usepackage{amsthm}
\usepackage{enumitem}
\usepackage{extarrows}
\usepackage{verbatim}
\usepackage{dsfont}
\usepackage{xcolor}
\usepackage{textcomp}
\usepackage{tikz-cd}
\usepackage{smartdiagram}
\usepackage{tensor}
\usepackage{mathtools}
\usepackage{float}
\usepackage[toc,page]{appendix}
\usepackage{enumitem}
\usetikzlibrary{decorations.pathmorphing}
\usetikzlibrary{backgrounds, chains, positioning, shadows}

\theoremstyle{definition}

\newtheorem{stp}{Step}

\makeatletter

\newcommand*{\rom}[1]{\expandafter\@slowromancap\romannumeral #1@}

\makeatletter
\begin{document} 
	\title{Lateral land movement prediction from GNSS position time series in a machine learning aided algorithm}   
	\author[M. Kiani]{M. Kiani}
	
	\thanks{Corresponding author, email: mostafakiani@alumni.ut.ac.ir, tel:+989100035865}
	\date{}
\begin{abstract}
We investigate the accuracy of conventional machine learning aided algorithms for the prediction of lateral land movement in an area using the precise position time series of permanent GNSS stations. The machine learning algorithms that are used are tantamount to the ones used in \cite{Kiani1}, except for the radial basis functions, i.e. multilayer perceptron, Bayesian neural network, Gaussian processes, k-nearest neighbor, generalized regression neural network, classification and regression trees, and support vector regression. A comparative analysis is presented in which the accuracy level of the mentioned machine learning methods is checked against each other. It is shown that the most accurate method for both of the components of the time series is the Gaussian processes, achieving up to 9.5 centimeters in accuracy.
\end{abstract}
\maketitle
$Key words:$ GNSS position time series, machine learning, lateral land movement prediction
\section{Introduction}
Machine learning prediction algorithms are increasingly gaining attention in the field of geoscience \cite{Kiani1}, \cite{Kiani2}. These methods are different from the traditional, statistical methods for approximation and interpolation \cite{Kiani3}-\cite{Kiani11}. The difference lies in the fact that machine learning methods are based on training and then extrapolation, whereas the interpolation and classical geodetic approximation problems use the concept of training and then finding the values at the queried points, which lie in the boundary defined by the training data. 

Machine learning algorithms are even more powerful than the statistical methods such as Theta \cite{Assimakopoulos} for prediction purposes \cite{Kiani1}, \cite{Kiani2}. In \cite{Kiani1}, an analysis for the efficiency of the Generalized Regression Neural Networks (GRNN) for the prediction of GNSS time series values is presented. The study is focused on many different aspects of the prediction error, such as the availability of gaps, different weather conditions, and the accuracy of the observations in the training process. In \cite{Kiani2}, an algorithm is presented for the vertical land movement. The high level of precision that is achieved in this paper has motivated us to investigate a new algorithm for the horizontal land movement prediction. It is interesting to see the differences between the results of \cite{Kiani2} by applying the algorithm to be proposed to the same data in \cite{Kiani2}.

The rest of this paper is organized as follows. In section 2 the algorithm is proposed. In section 3 the numerical results are presented. In the end, section 4 is devoted to conclusions.       
\section{How the algorithm works}
In contrast to the precise computations that were needed for the calculations of the vertical land movements \cite{Kiani1}, including tidal and atmospheric effects, the lateral land movement prediction does not need such computations. This is because mostly the vertical components are affected by the mentioned forces \cite{Jin}. Hence, the following algorithm is proposed.

\begin{stp}\label{stp1}
The first thing to do is to consider the number of training data and choose the performance criteria. Let us denote the time series values by $p_k$, which are the sampled values of a functions at times $t_k$ ($k=1,2,...,n$). The maximum number of data we have at our disposal is $m$. As \cite{Kiani1} and \cite{Kiani2} note, one can use the Root of Mean Squared Errors (RMSE), a good indicator of precision for both the training and prediction phases. Hence, in this paper RMSE is used as the training performance criteria. 
\end{stp}
\begin{stp}\label{stp2}
The second step is to predict the next values of the time series based on the model obtained by training in the previous step. In this phase, the conventional machine learning methods are used. These are Multi-Layer Perceptron (MLP), Bayesian Neural Network (BNN), Gaussian Processes (GP), K-Nearest Neighbor (KNN), GRNN, Classification And Regression Trees (CART), and Support Vector Regression (SVR). For more information on these methods refer to \cite{Ahmed}, \cite{Makridakis}, and \cite{Alpaydin}. It is essential to notice that in this step, the times of the acquisition of the training data are normalized, i.e. we have
\begin{equation}\label{eqn1}
\overline{t}_k=\frac{t_k-t_1}{t_m-t_1},\quad k=1,...,m,
\end{equation}  
where $\overline{t}_k$ denotes the normalized time.      
\end{stp}
\begin{stp}\label{stp3}
The final step is the assessment of the prediction accuracy. This is done by three reliable indices of prediction performance assessment, which are RMSE, Mean Absolute Scaled Error (MASE), and Mean of Absolute Errors (MAE). For the definition of these indices the reader is referred to \cite{Kiani1}, \cite{Kiani2}, \cite{Ahmed}, and \cite{Makridakis}. 
\end{stp}
The following diagram summarizes the steps involved in the algorithm. 
\begin{center}
	\resizebox{0.4\linewidth}{!}{%
		\begin{tikzpicture}[
		node distance = 10mm and 0mm,
		start chain = going below,
		box/.style = {rectangle, rounded corners, draw=gray, very thick,
			minimum height=16mm, text width=60mm, align=flush center,
			top color=#1!90, bottom color=#1!90,
			on chain},
		down arrow/.style = {
			single arrow, draw,
			minimum height=2.5em,
			transform shape,
			rotate=-90,
		}
		]
		\node (n1) [box=cyan]{Train the machine learning algorithm};
		\node (n2) [box=cyan]{Predict the next horizontal values of the time GNSS series};
		\node (n3) [box=cyan]{Assess the precision of the prediction};
		\draw [black, thick, ->] (n1) edge (n2)(n2) edge (n3);
		
		\end{tikzpicture}
	}
\end{center}

\section{Numerical results: a study for different times series in Europe}
\begin{figure}[H]
	\centering
	\includegraphics[width=1\linewidth]{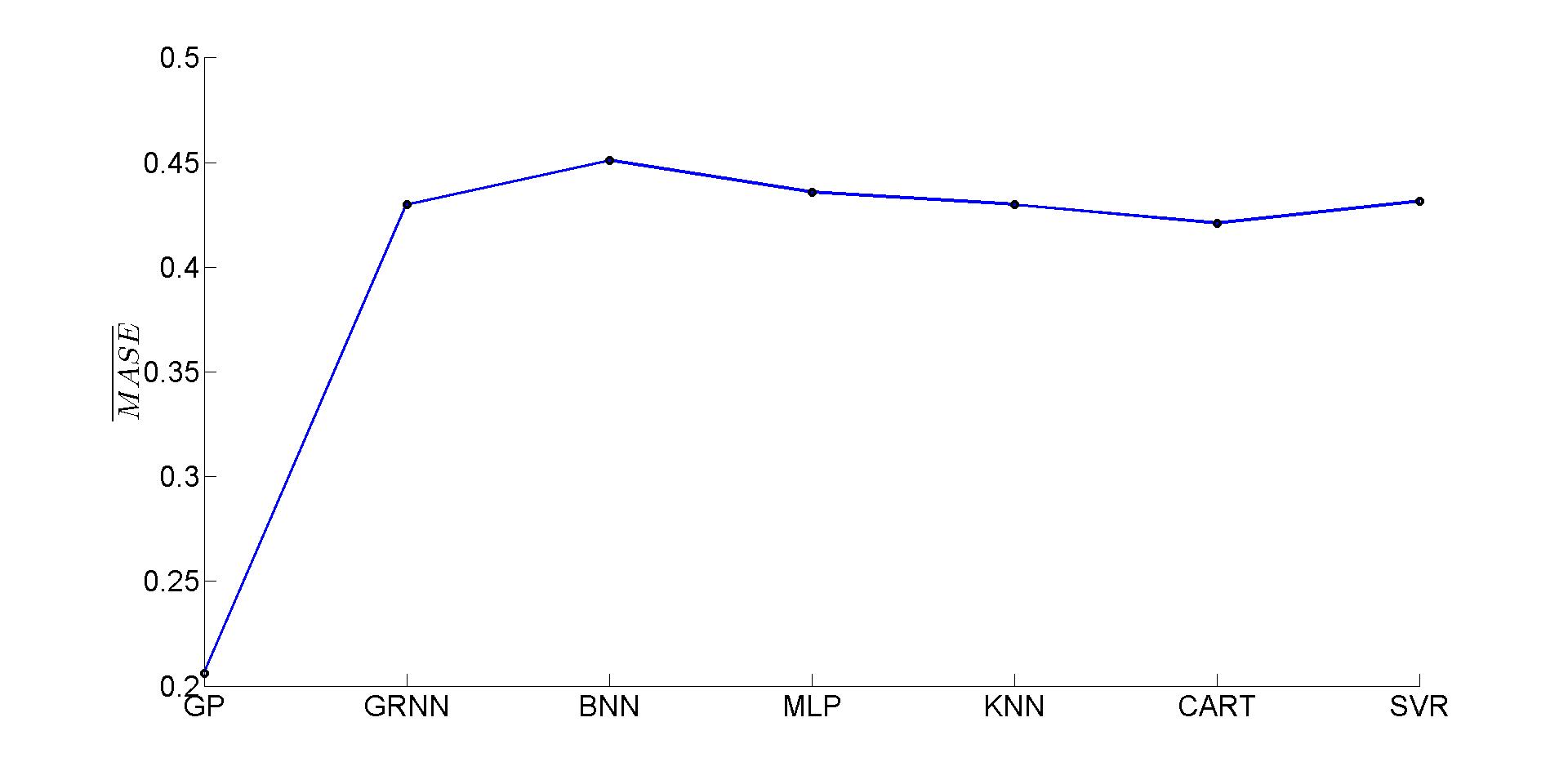}
	\caption{First component of the time series, $MASE$ criterion for the prediction performance assessment}
	\label{fig1}
\end{figure}

\begin{figure}[H]
	\centering
	\includegraphics[width=1\linewidth]{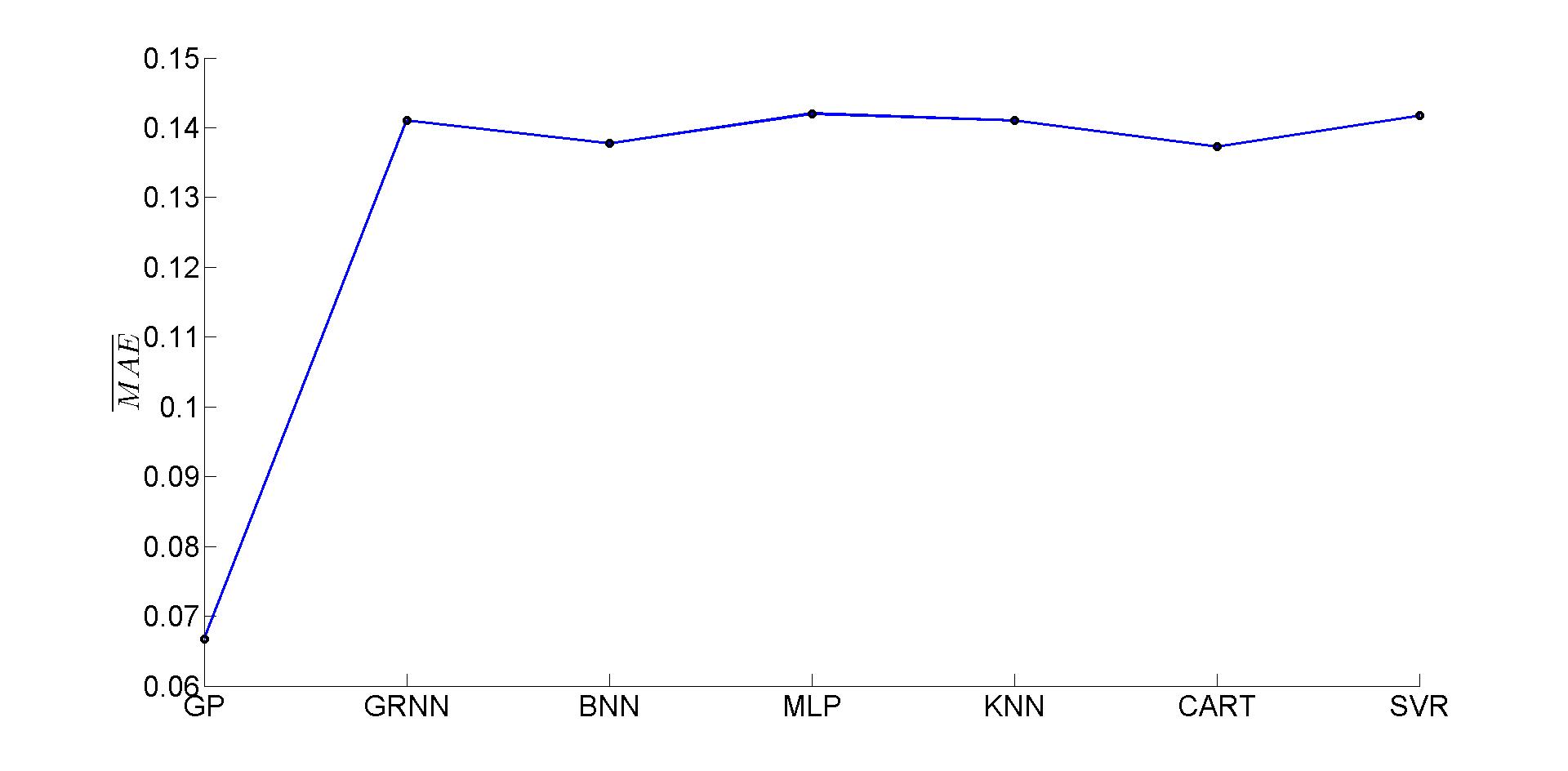}
	\caption{First component of the time series, $MAE$ criterion for the prediction performance assessment}
	\label{fig2}
\end{figure}

\begin{figure}[H]
	\centering
	\includegraphics[width=1\linewidth]{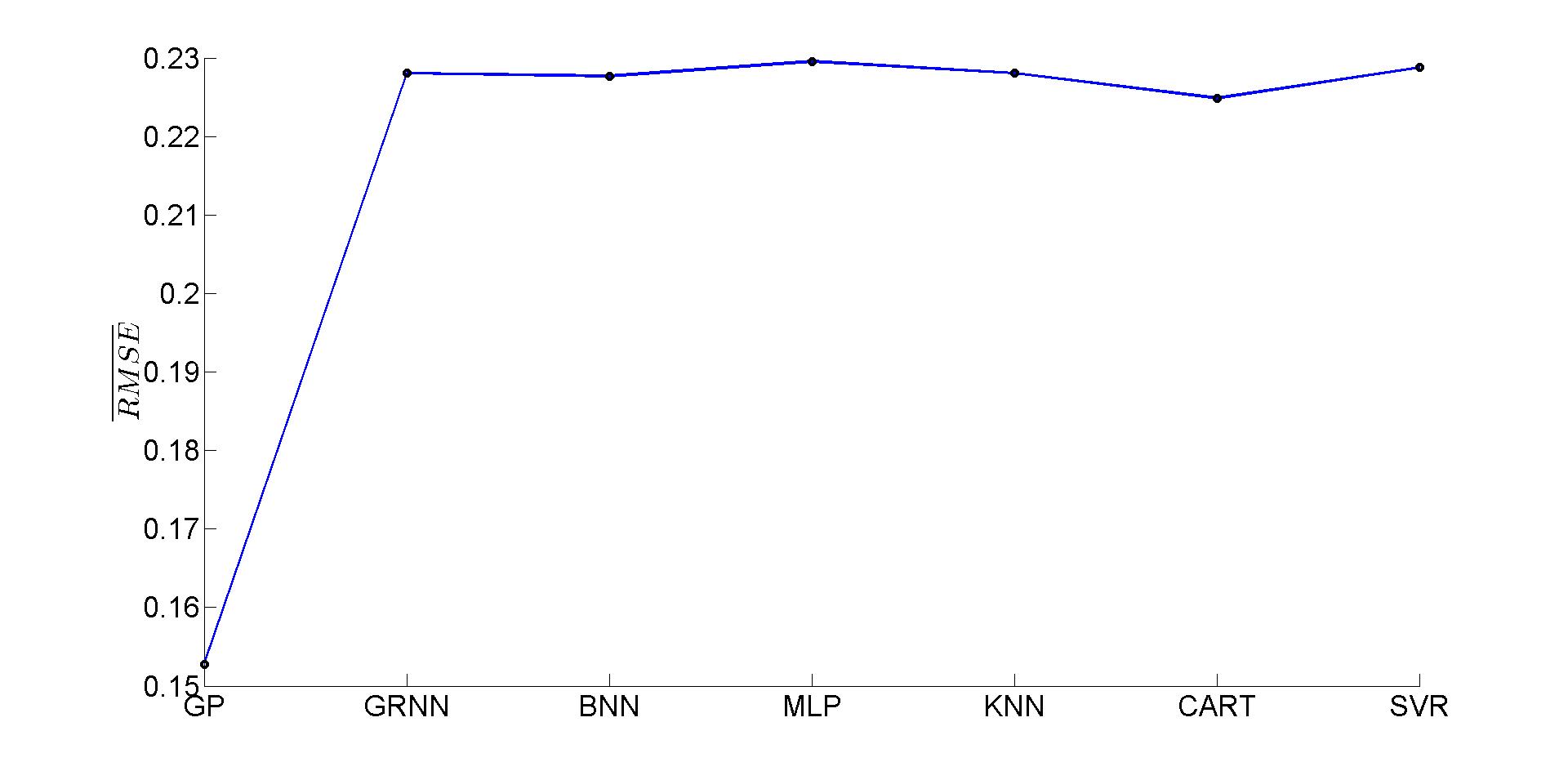}
	\caption{First component of the time series, $RMSE$ criterion for the prediction performance assessment}
	\label{fig3}
\end{figure}

\begin{figure}[H]
	\centering
	\includegraphics[width=1\linewidth]{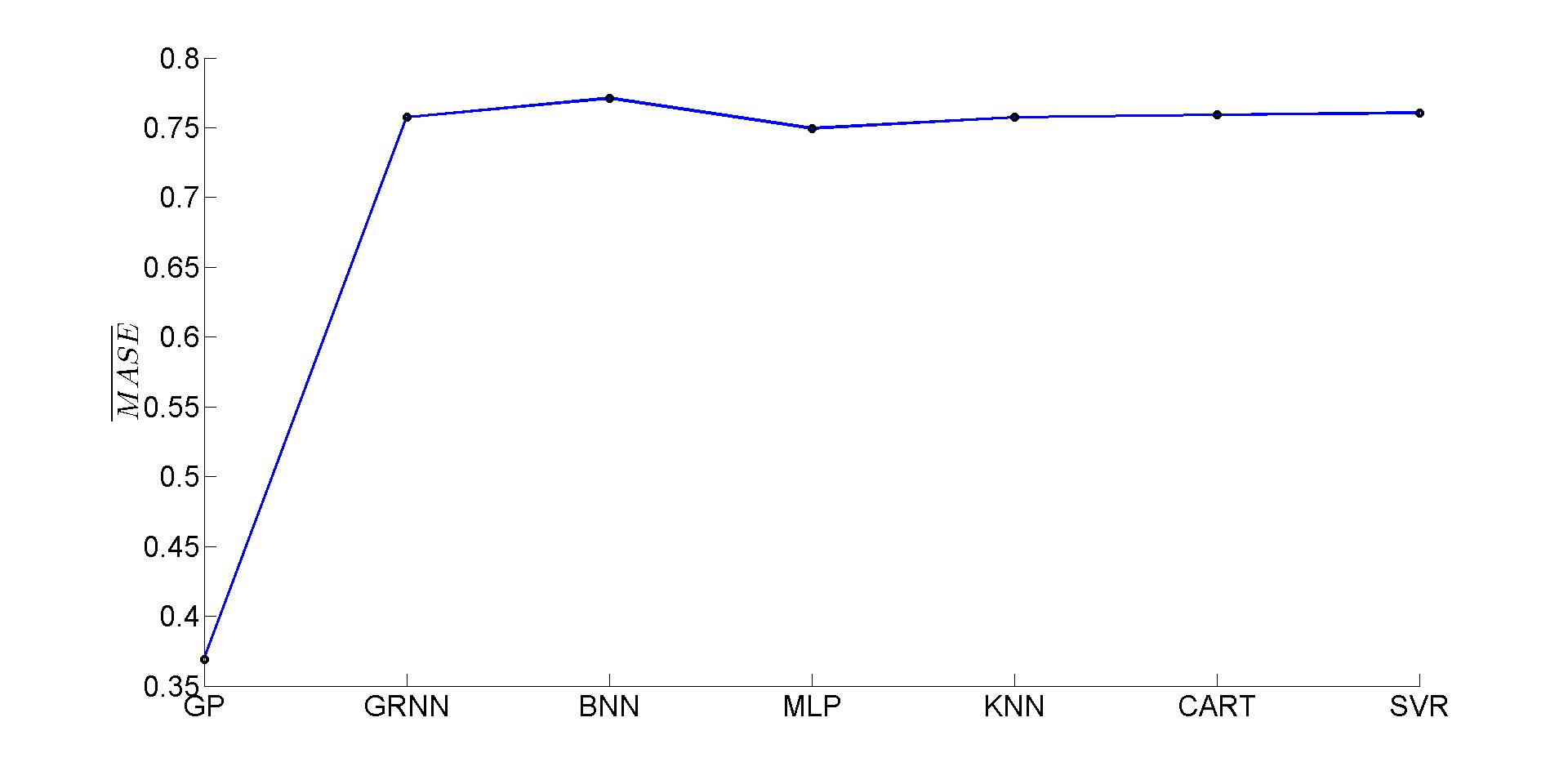}
	\caption{Second component of the time series, $MASE$ criterion for the prediction performance assessment}
	\label{fig4}
\end{figure}

\begin{figure}[H]
	\centering
	\includegraphics[width=1\linewidth]{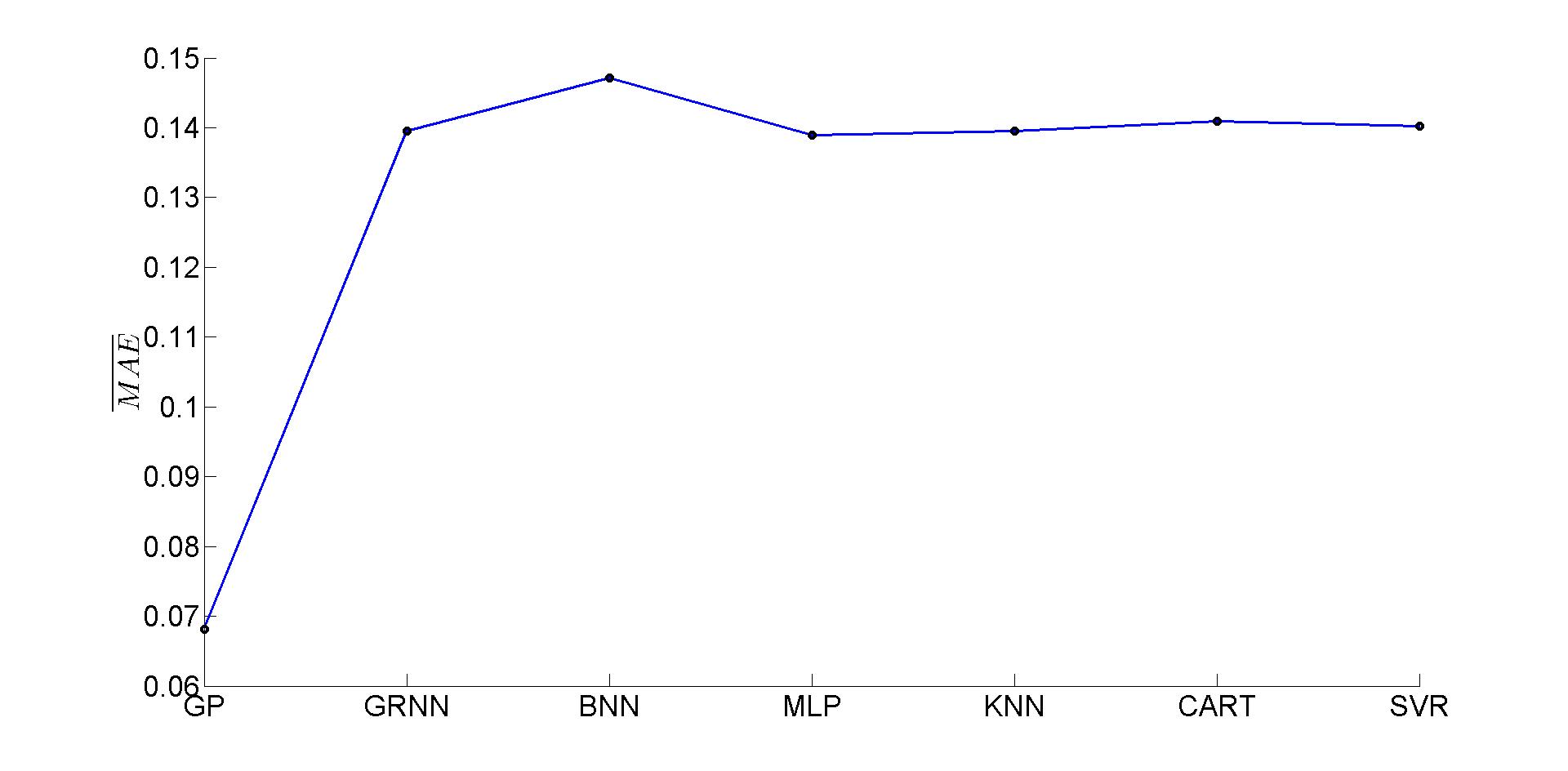}
	\caption{Second component of the time series, $MAE$ criterion for the prediction performance assessment}
	\label{fig5}
\end{figure}

\begin{figure}[H]
	\centering
	\includegraphics[width=1\linewidth]{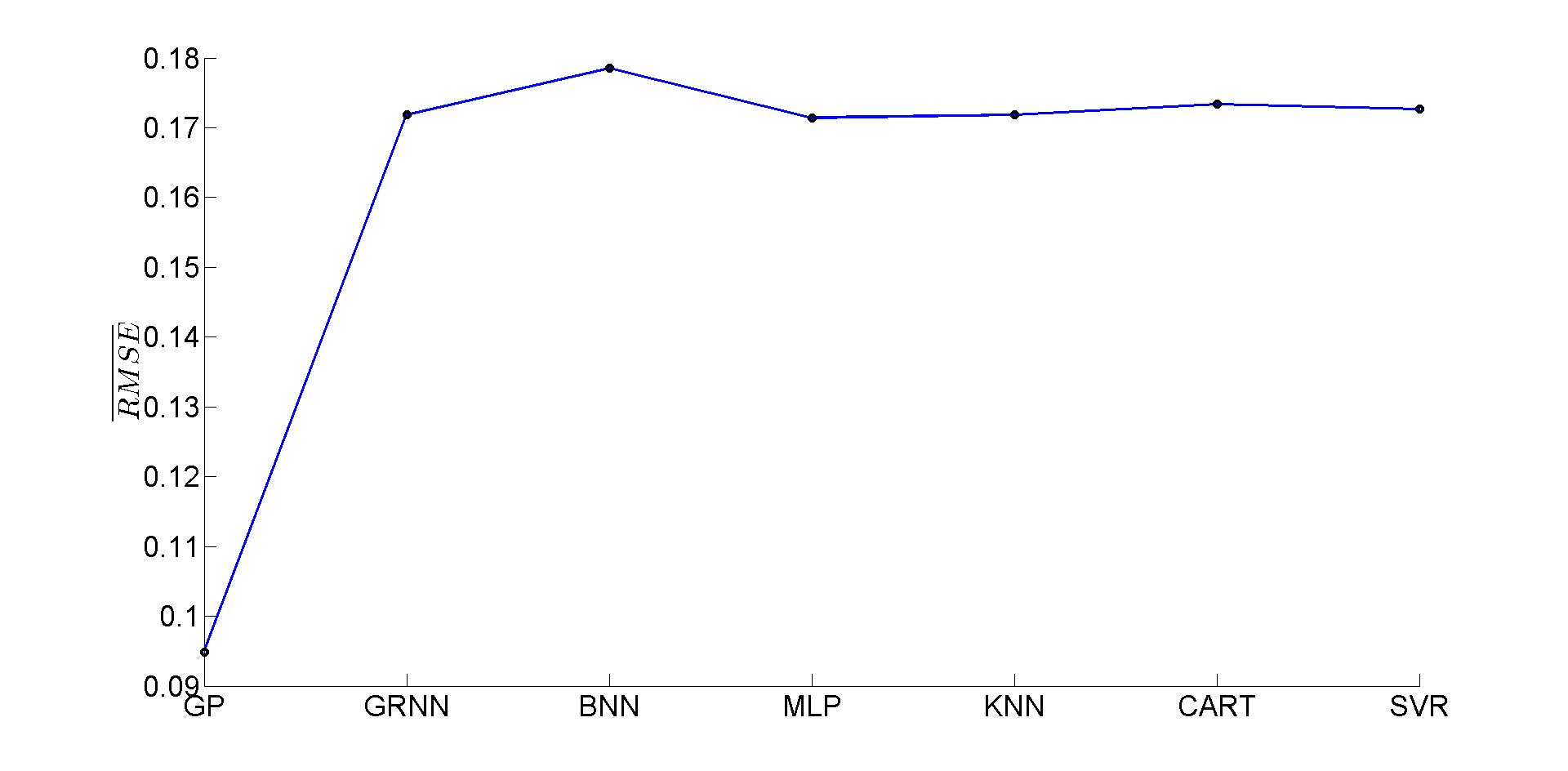}
	\caption{Second component of the time series, $RMSE$ criterion for the prediction performance assessment}
	\label{fig6}
\end{figure}

In this section, the simple algorithm mentioned in the previous section is used for the same time series used in \cite{Kiani1} and \cite{Kiani2}. The data are taken from \cite{Blewitt}. The following figures show the prediction performance of the machine learning algorithms. Note that the first and second component of the time series are treated separately. By the first component we mean the $X$ coordinate of the point in the IGS14 system. The $Y$ component of the point in the IGS14 system is the equivalent of the term "second component of the time series". The predicted values are used to compute the $MASE$, $MAE$ and $RMSE$ in each of the 14 stations. Then the averages of all the 14 $MASE$, 14 $MAE$ and 14 $RMSE$ are calculated and denoted by $\overline{MASE}$, $\overline{MAE}$ and $\overline{RMSE}$, respectively. The figures \ref{fig1}-\ref{fig6} represent $\overline{MASE}$, $\overline{MAE}$ and $\overline{RMSE}$ for the first and second component of the time series.
As it can be understood from the figures above, in both of the first and second components of the time series the GP method works the best. In the time series data used in this study the first component has a lower accuracy than the second one, which renders an important point about the nature of the machine learning methods: they are dependent on the precision of the training data; the more accurate the training data the better the performance of the time series prediction is. Note, however, the first component of the item series is less winding than the second component, which will also give the idea that the machine learning methods have a better performance if the time series is not much variable; the less variable the values of the time series are, the more accurate the prediction is.
\section{Conclusion}
A simple machine learning aided algorithm is presented by which the horizontal components of the GNSS position time series can be accurately predicted. The most accurate machine learning algorithm is the Gaussian processes, which can achieve up to 9.5 centimeters in accuracy. 

The promising results in this paper should motivate the researchers in the field of geoscience to work on the machine learning algorithms more.  
    
\end{document}